%% file: main.tex
\documentclass[manuscript]{aastex631}

\newcommand{\Msol}[1]{$#1 \, M_\odot$}

\graphicspath{{./}{figures/}}

\received{2022 October 18}
\revised{2023 January 12}
\accepted{2023 January 13}
\submitjournal{MNRAS}

\shorttitle{Helium-rich MWDs}
\shortauthors{Hardy et al.}

\begin{document}

\title{Spectrophotometric analysis of magnetic white dwarf II: Helium-rich compositions}

\author{François Hardy}
\affiliation{Département de Physique, Université de Montréal, Montréal, Québec H3C 3J7, Canada}

\author{Patrick Dufour}
\affiliation{Département de Physique, Université de Montréal, Montréal, Québec H3C 3J7, Canada}

\author{Stefan Jordan}
\affiliation{Astronomisches Rechen-Institut am Zentrum für Astronomie, Universität Heidelberg, Germany}

\begin{abstract}
    We present an analysis of all single white dwarf stars known to exhibit spectroscopic signatures of neutral helium line splitting due to the presence of a strong magnetic field.
    Using state-of-the-art models taking into account the effects of magnetic fields on the synthetic spectra, we determine effective temperatures, surface gravities and masses for the stars in our sample.
    Our analysis uses data from the second and third Gaia (early) data release, photometric data from diverse surveys such as the Sloan Digital Sky Survey and Pan-STARRS, and archived spectroscopic data.
    We are able to sucessfully reproduce the spectra of 8 objects using an offset dipole geometry while several others seem to require either a more complexe geometry or a different chemical composition.
    We also highlight a group of hot featureless white dwarfs that are most probably highly magnetic objects whose spectra are completely smeared due to the field strength distribution across the surface.

\end{abstract}

\keywords{stars: white dwarfs --- stars: magnetic field --- techniques: photometric --- techniques: spectroscopic}


\section{Introduction} \label{s:intro}


Magnetism is found to be present in about 10 to 20\% of white dwarfs \citep{Kawka2007, Giammichele2012, Bagnulo2021}.
According to the compilation of \citet{Ferrario2015a} and from the Montreal White Dwarf Database \citep[MWDD,][]{MWDD}, the vast majority of magnetic white dwarfs (MWDs) identified in the literature are found to have a hydrogen-rich atmosphere.
Such white dwarfs have been examined in detail in the first paper of this series \citep[][hereafter Paper I]{Hardy2022}.
On the other hand, magnetic white dwarfs with elements other than hydrogen as their main constituent appear to be much rarer.
Magnetism has been detected in a handful of cool helium-rich objects such as DQs \citep{Schmidt1995, Schmidt1999, Vornanen2010, Vornanen2013} and DZs \citep{Reid2001, Dufour2006, Hollands2015}.
Magnetism is also detected through circular spectropolarimetry in a number of featureless cool DC white dwarfs, some of which are also probably helium-dominated \citep[][]{Putney1997}.
On the hotter side, a large fraction of the carbon-oxygen dominated atmosphere white dwarfs, the so-called "Hot DQ", also exhibit traces of magnetism \citep{Dufour2008, Dufour2010}.
In the DB temperature range (roughly 12,000 K to 30,000 K), however, the incidence of magnetism appear to be quite rare.
While a very weak magnetic field (1.3 kG) has been detected through asteroseismology in GD358 \citep[][]{Winget1994}, unambiguous detection of helium in a magnetic white dwarf was only first confirmed with detailed modeling by \citet{Achilleos1992} for the star Feige 7 ($\sim$ 35 MG).
A weaker magnetic field ($\sim$ 100 kG) has also been detected from circular polarization in the DBA white dwarf LB 8827 \citep[][]{Wesemael2001}. More recently, \citet{Richer2019} announced the discovery of a magnetic DB star in the M47 cluster.
This star shows magnetic splitting of helium lines coherent with a field between 2 and 3 MG, but also features that indicate it may be a binary system (MWD + late-type companion). Four magnetic DBs have also been identified in the course of the Hamburgh ESO survey by \citet{Reimers1998}, although no formal analysis has ever been attempted to our knowledge.
Finally, from comparison of observed features with Zeeman splitting calculations of helium lines, three more magnetic DB, GD 229, HE 1211$-$1707, and HE 1043$-$0502 (they are included in our sample under the names J2012+3113, J1214$-$1724 and J1046$-$0518 respectively), have been identified \citep[][]{Jordan1998, Wickramasinghe2002}.
Detailed modeling similar to that presented for magnetic DAs in Paper I, however, has never been attempted for all these objects.

In Paper I, we examined 651 objects that had previously been classified as magnetic hydrogen-rich white dwarfs in the literature.
During this process, we found 38 objects that we could not fit within our assumed theoretical framework, suggesting that some of these could have a non-hydrogen-rich composition.
Furthermore, the MWDD lists numerous white dwarfs classified as DBH that, to our knowledge, have never been analyzed properly using magnetic model atmosphere.
We thus aim here to visit/revisit all those potentially helium-rich objects in light of the mass availability of precise trigonometric parallax measurements \citep[][]{Gaia2016,Gaia2018,Gaia2021} and broadband photometry \citep[][]{York2000, Chambers2016}.
However, since our sample consists mostly of spectra with low signal-to-noise ratio and medium resolution from the Sloan Digital Sky Survey (see below), we restrict ourselves to objects in the DB temperature range ($\sim$ 12,000-30,000 K) that should show detectable signs of helium-line splitting when highly magnetic ($B \ge 2 MG $).
This paper aims to model those objects with the assumption of a helium-rich composition and offset dipole geometry, leaving other compositions and geometries to future studies.
In \autoref{s:observations}, we present our sample, in \autoref{s:theory} we describe our theoretical framework and fitting method.
\autoref{s:results} present our detailed analysis and assessment for each star and we conclude in \autoref{s:conclu}.

\section{Observations} \label{s:observations}

\subsection{Sample Selection} \label{ss:sample}

We aim to analyze in a homogeneous fashion all known DB magnetic white dwarfs that have been identified in the literature over the years \citep[MWDD,][and references therein]{Ferrario2015a, Ferrario2020}.
We also include the many objects that have been flagged as DBH by visual inspection of thousands of SDSS spectra \citep[32 stars, mostly from][]{Kleinman2013,Kepler2015, Kepler2016} as well as all the objects that we were unable to fit using hydrogen-rich models in Paper I (38 stars).
This brings the total number of potentially magnetic DB white dwarfs in our sample to 83 stars when removing duplicates found in more than one of the above categories.
However, optical spectra were not available for 4 objects, leaving us with a total sample of 79 stars to examine (see Table 1).
Apart from those mentioned in the introduction, these white dwarfs have never, to our knowledge, been subject to a detailed analysis using magnetic synthetic spectra.

\input{names}

In order to have an analysis as homogeneous as possible, we use in priority SDSS \citep{York2000} \emph{ugriz} point-spread function magnitudes when available. If \emph{ugriz} data was not available for an object, we relied on Pan-STARRS \citep{Chambers2016} \emph{grizy} or previously published \emph{JHK} and \emph{BVRI} photometry, in that order of priority.
Of our 79 objects, 13 ($\sim 16$\%) had no \emph{ugriz} data.
For those situations, we had to use \emph{grizy} data for 10 stars, \emph{BVRI} photometry for 2 stars and \emph{B + JHK} for a single star.

The majority (67 out of 79; 84\%) of the spectra used for this work are collected from SDSS.
The spectra of the 12 objects without SDSS spectroscopy were taken from
\citet{Kilic2020, Bergeron1997, Bergeron2001, Giammichele2012, Limoges2015}
or archival data obtained by the Montreal group over the years.
Spectra for the stars presented in \citet{Wickramasinghe2002} (GD 229, HE 1211$-$1707, and HE 1043$-$0502) were kindly made available to us by L. Ferrario (private communication).

\section{Theoretical Framework} \label{s:theory}

The theoretical framework and fitting method used in this study are very similar to that described in Paper I, except that a pure helium composition is assumed. Below we briefly review the methodology, emphasizing the differences with Paper I.

The first detailed investigation of magnetic white dwarfs with helium absorption features, that of \citet{Achilleos1992}, relied on line data calculation from \citet{Kemic1974}.
Those first order calculations, however, are only valid up to about 20 MG and tables quickly need to be extrapolated outside their range of validity for most practical purposes.
Subsequently, more detailed calculation performed in Heidelberg provided precise data for a large number of energy states in the 300-700 MG range that allowed the identification of stationary neutral helium lines \citep[][]{Jordan1998}, but the full parameter space has yet to be explored systematically using those data.

\begin{figure}
    \centering
    \includegraphics[width=0.7\linewidth]{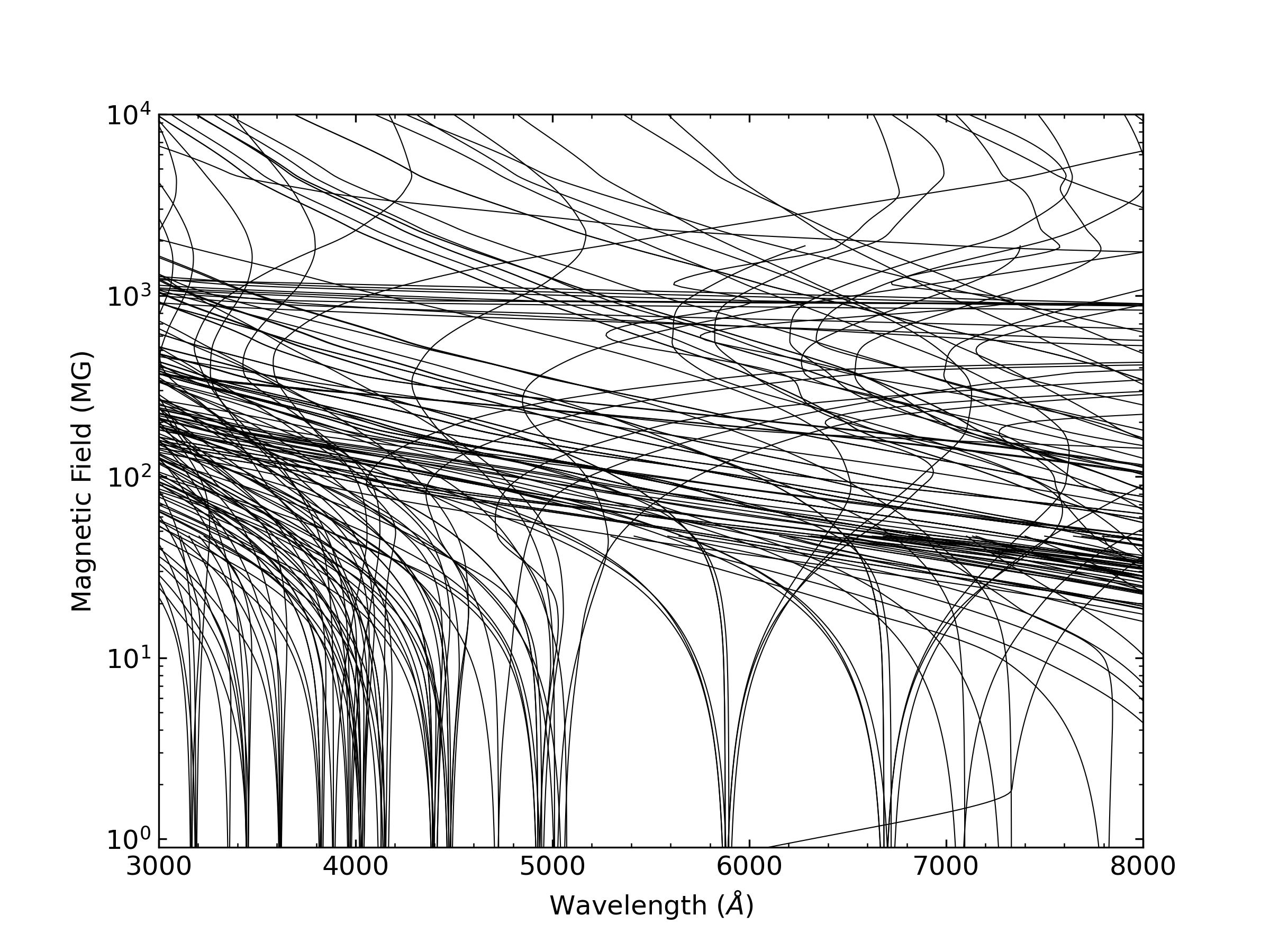}
    \caption{\label{f:helium} Position of helium lines between 3000 and 8000 \AA{} with increasing magnetic field. We show only lines with zero-field wavelengths below 12 000 \AA{} for clarity.}
\end{figure}

\begin{figure}
    \centering
    \includegraphics[width=0.4\linewidth]{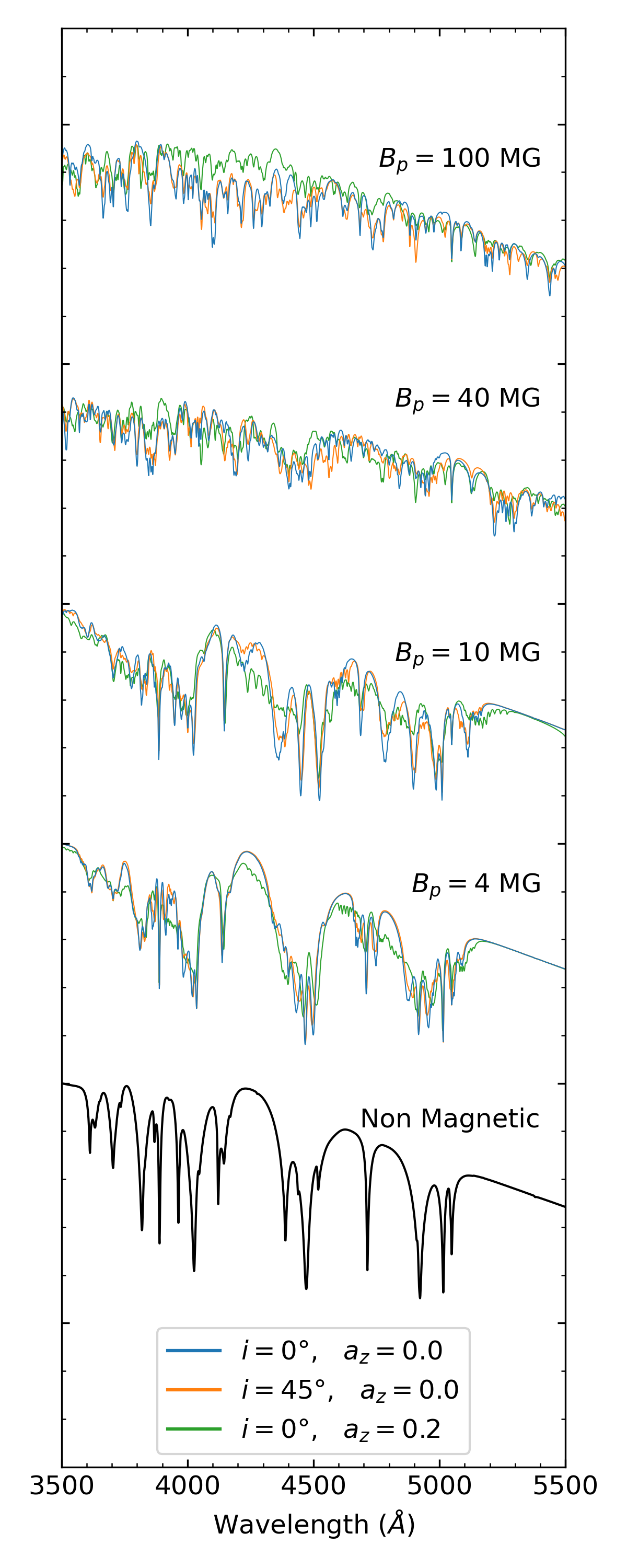}
    \caption{\label{f:repr_synth} Representative reconstructed synthetic spectra of magnetic DB white dwarfs at various field strength, inclination and offsets for an effective temperature of 20,000 K and a surface gravity of $\log{g}=8.0$. The spectra are normalized by the flux value of the continuum at 3500 \AA{} and offset by an arbitrary factor for clarity.}
\end{figure}

In this work we use the helium line splitting data described in \citet{Jordan2001}, \citet{Becken1999} and \citet{Becken2000}.
The data contained transitional energies, wavelengths, and various quantum quantities needed to compute oscillator strengths, for a wide range of magnetic field intensities from 0 to over 200 GG ($200 \times 10^3$ MG), for each of those components.
We implemented an interpolating scheme developed by \citet{Steffen1990} to obtain the component data for arbitrary field intensities, with a linear extrapolation when required.
This extrapolation occurs at over 200 GG for most components, although some weaker components only have calculations up to 500 MG.
As most stars in our sample have a dipolar magnetic field $B_p$ of under 100 MG with only a very limited number of stars with $B_p$ over 500 MG, this extrapolation is not expected to be frequent.
A wide subset of the line data is illustrated in \autoref{f:helium}, where the line position in relation to the applied magnetic field can be seen.
For magnetic field strength above about 5 MG, large non-linearities become very important with many components shifting rapidly by hundreds of Angstrom as the field strength increases.
As a result, it becomes difficult to recognize the helium signatures in the most magnetic objects.
Note that for this exploratory work, we simply assume that the broadening treatment for each line component is the same as for non-magnetic white dwarfs as no broadening calculations in the presence of strong magnetic field have been performed to our knowledge.
Strongly magnetic white dwarf spectra may thus not be accurately be represented within our theoretical framework.
\autoref{f:repr_synth} shows typical helium-rich synthetic spectra assuming a dipolar geometry for various field strength, inclination and dipole offset.
For some geometries at the highest field intensities, the spread of magnetic field strength over the surface leads to absorption features spread all over large wavelength range, making lines almost unrecognizable.
In some cases, the density of lines can almost mimic a featureless DC spectrum while it could easily be interpreted as noise in others.
As we will see below, some of these configurations may explain hot stars with spectra that are globally featureless but should normally show strong hydrogen/helium lines if non-magnetic, assuming the effective temperature inferred by their photometric energy distribution.
We believe these "Hot DC" are probably highly magnetic white dwarfs with line features that are so spread out that almost no spectral features can be recognized anymore.

The methodology to obtain the atmospheric parameters is the same as in Paper I.
Briefly, photometry is first fit using multiple model grids with varying magnetic field intensities, and the resulting atmospheric parameters are then used to determine the dipolar geometry that best reproduces the spectroscopic observations.
The magnetic field geometry is reconstructed from pre-calculated grid (13 GB of data) covering 5 parameters: $\log{g}=7.0$ to $9.5$ (in steps of 0.5 dex), $T_{\rm{eff}}=$12,000 K to 30,000 K (in steps of 3,000 K), field intensity (0 to 500 MG in 54 log steps), the angle between the line of sight and the normal of the surface ($\psi$ from 0 to $\pi$ in 6 steps), and the limb darkening ($\mu = \cos{\theta}$ from 0 to 1 in 4 steps).
While this exploratory work only considers offset dipole geometries, our technique allows us to rapidly construct any desired geometry.

\section{Results} \label{s:results}

Using the fitting technique described in detail in Paper I, we fitted all 79 objects in our sample in a homogeneous way assuming single DB synthetic spectra with a dipolar geometry.
Each solution was then carefully inspected by eye to judge the quality of the fit.
From this exercise, we found that, similarly to many DAH from Paper I, the DBH classification for 18 objects did not hold up against detailed comparison with magnetic synthetic spectra.
The alleged splitting observed in some lines was probably, in hindsight, simply noise.
While better spectroscopic observation may reveal that some of these stars are magnetic after all, we believe classifying these objects as magnetic based on these low signal-to-noise ratio spectra was perhaps a bit premature.
We thus reject these objects from our analysis and reclassify them as DB until proof to the contrary is obtained.
Five other objects were also rejected since the spectra were too noisy to assess their magnetic nature.
Furthermore, fit of the energy distribution for 20 objects indicated an effective temperature below $\sim$12,000K, too cool to show helium features.
Again, noise in the spectra was probably interpreted as a sign of magnetism for most of these. Some of these objects may turn to genuinely be magnetic white dwarfs when spectropolarimetric observations become available, but no such assessment could be made based on the spectra at hand.
The list of white dwarfs previously classified as magnetic that we reject for our analysis is presented in \autoref{t:rejected}.

\input{rejected_table}

From our fit assuming offset dipole geometry, we could only firmly confirm magnetism in 8 DB stars in our sample (Figures \ref{f:good_fit_0} to \ref{f:good_fit_7}).
With the exception of J0043-1000 \citep[Feige 7][]{Achilleos1992} these are, to our knowledge, the very first magnetic DB to be modeled with detailed helium line splitting calculations.
Dipole solutions appear to be an excellent representation for those stars except for J0043-1000 and J0856+1611.
In the first case, the small discrepancies that are observed can be explained by the fact that our grid does not include small traces of hydrogen that are known to be present (hints of H$\alpha$ can also be observed in J0305+3747 and J2323-0046).
Furthermore, \citet{Achilleos1992} also showed that time resolved spectroscopy revealed composition changes across the surface as the star rotates, something that is not captured by the single spectra used in this study.
In the case of J2323-0046, the predicted depth of the helium lines does not agree with the observations.
This is reminiscent of the shallow features found for numerous magnetic DA in Paper I, although the differences are more subtle here. It is also possible magnetic spots on the surface produce temperature variations that may explain why the lines are not as deep as expected from its photometric atmospheric parameters.
\autoref{t:results} present the atmospheric parameters for the confirmed magnetic DB stars in our sample.
We note that two objects have surface gravities lower than $\log{g}$=7, the lowest value in our grid.
Although this is a very small number of individuals, the average mass is around \input{average_mass}.
It thus appears that magnetic helium rich white dwarfs can be found at low, normal and high mass range, offering little indication of preferred formation channel.

\input{figures/mwd_db.tex}

\input{results_table}

\begin{figure}
    \centering
    \includegraphics[width=\linewidth]{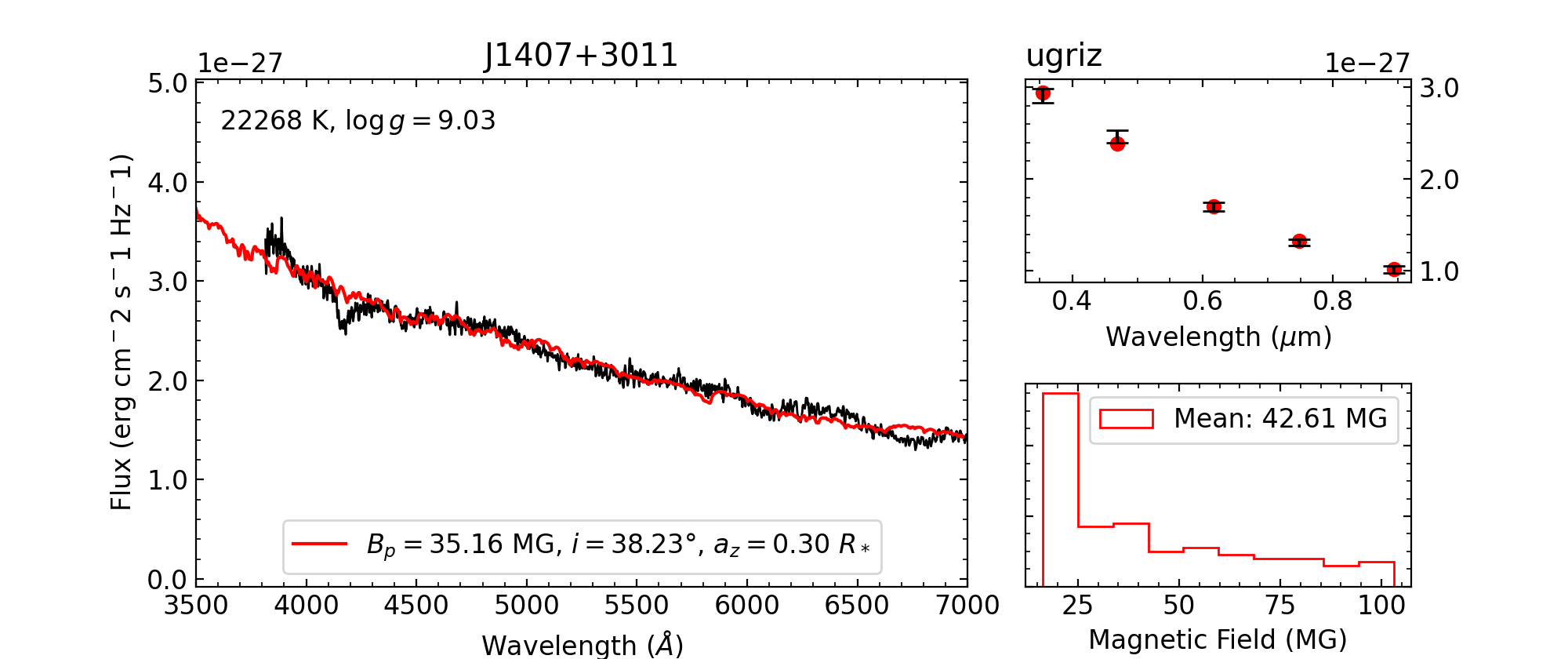}
    \caption{\label{f:4200A} Star J1407+3011 (SDSS J140750.65+301130.2) where the spectrum is mostly featureless, with the exception of one unidentified (and not reproduced) feature near 4200 \AA{}.}
\end{figure}

\begin{figure}
    \centering
    \includegraphics[width=\linewidth]{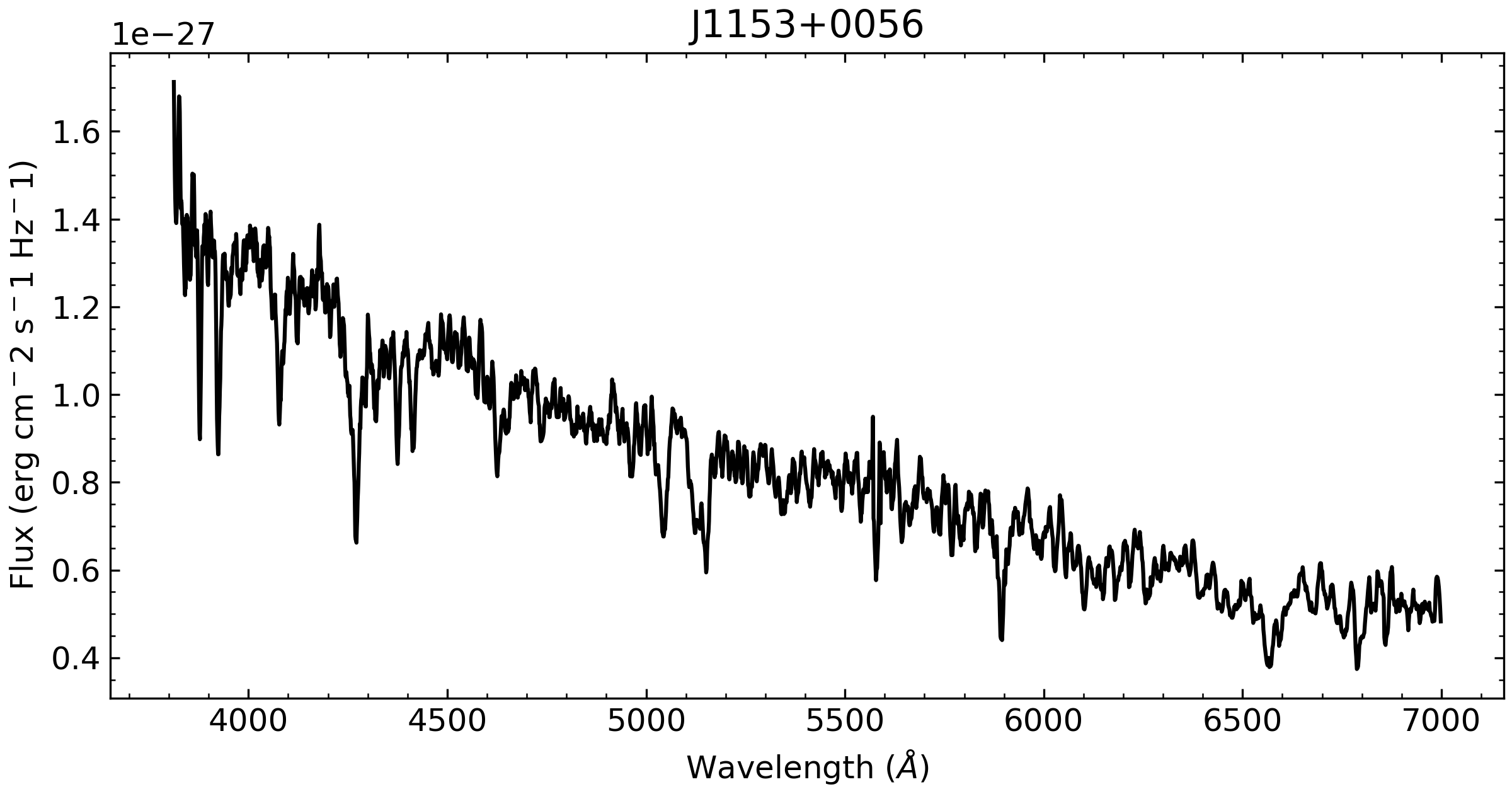}
    \caption{\label{f:hotdq} An example of carbon/oxygen dominated atmosphere (Hot DQ) with a strong feature near 4200\AA{}, J1153+0056 \citep[WD 1150+012,][]{Dufour2008}.}
\end{figure}

The rest of our sample comprises stars that we are unable to model using our magnetic helium models.
In fact, with the exception of the confirmed magnetic DB with a relatively small field strength J0212+0644 (\autoref{f:good_fit_4}), none of the stars that we were unable to fit (or that we tentatively labeled as Magnetic non DA) in Paper I could successfully be reproduced using the current theoretical framework.
Likewise, we were also unsuccessful in modeling the few objects that are known to exhibit spectroscopic signatures (stationary components) indicative of neutral helium in very high magnetic field \citep[][]{Jordan1998, Wickramasinghe2002}, GD 229, HE 1211$-$1707, and HE 1043$-$0502.
These objects possibly harbor extremely strong magnetic fields that cannot be simply reproduced assuming a simple offset dipole and other geometry will need to be explored in future studies.
Finally, we find several additional objects (see \autoref{t:nonDB}) showing a number of features that could also not be reproduced with either the hydrogen or helium grid that we computed.
While our work ruled out a simple dipole geometry for both hydrogen and helium-rich composition, it is possible that a much more complex field structure is at play.
Another possibility is that some of the absorption features are from another element than hydrogen and helium, carbon and oxygen being the prime suspects.
\citet{Dufour2008, Dufour2010} has shown that a large fraction of the so-called Hot DQ (see \autoref{f:hotdq} for an example spectrum) are magnetic at some level (a few hundred kG to 2-3MG).
Similar objects but with much stronger fields could be very difficult to identify in the absence of detailed calculation for the position of the various transitions as a function of magnetic field strength.
It is interesting to note, for instance, that many objects show a significant absorption feature around 4200 \AA{} (for example \autoref{f:4200A}).
It is tempting to attribute this to carbon as Hot DQ white dwarf tend to show a similar strong feature but at a slightly longer wavelength (see \autoref{f:hotdq}).
This hypothesis would necessitate that all other features would be smeared out.
However, in the absence of calculation of carbon/oxygen line position in very large magnetic fields, we refrain from drawing any conclusion at this point.

\input{magnetic_non_DB_table}

\begin{figure}
    \centering
    \includegraphics[width=\linewidth]{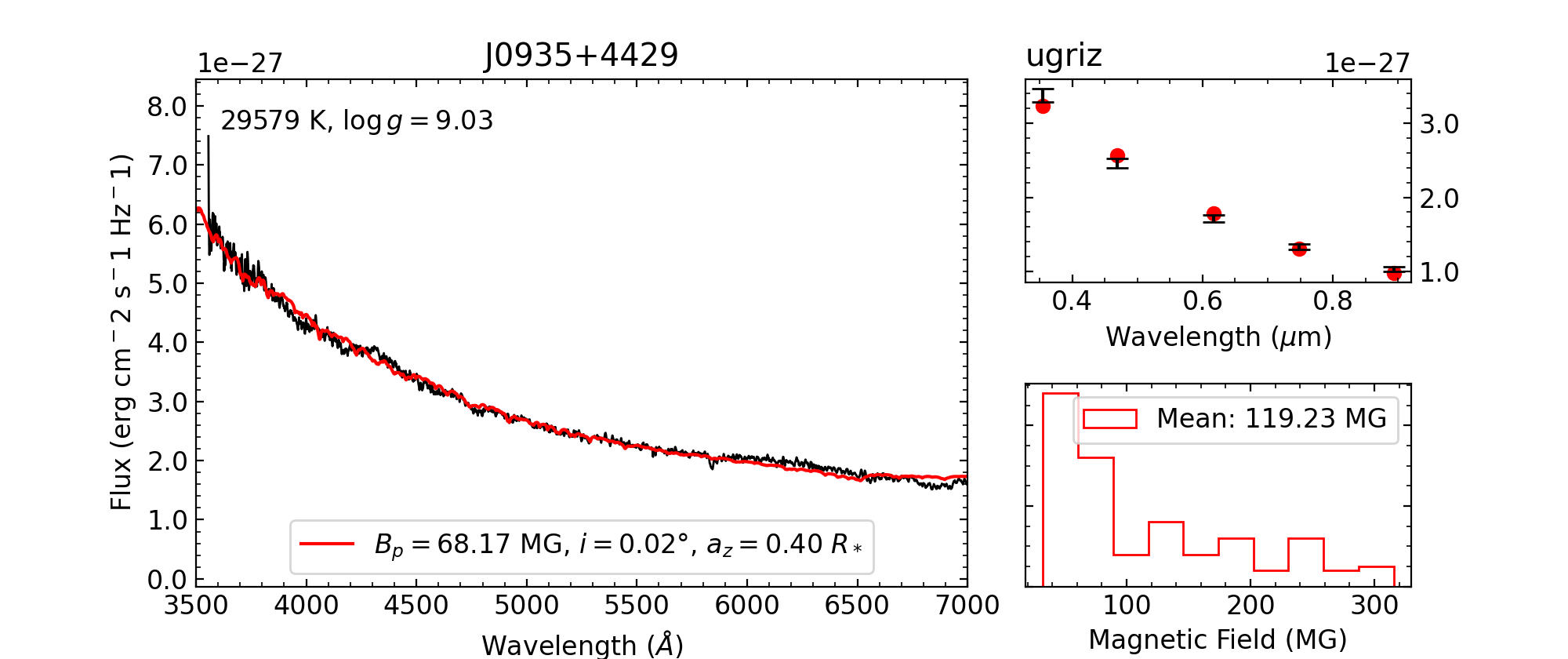}
    \caption{\label{f:hotdc} Star J0935+4429 (US 736) where the spectrum is almost featureless, and the offset dipole solution leads to minimal spectral features.}
\end{figure}

Finally, we note the presence of several objects that show almost continuous spectra.
This is very unusual, as both DA and DB white dwarfs should show strong lines in the optical at their inferred effective temperature (see \autoref{t:hotdc}).
It is, of course, difficult if not impossible to infer their chemical composition from spectrometry in these cases. We believe, however, that these "Hot DC" have geometries that produce surface field strength variations that result in spectral components almost completely smeared across the spectra.
\autoref{f:hotdc} shows an example of an offset dipole solution that leads to practically no spectral features.
It is important to note that, while the magnetic field geometries do reproduce the observations to an extent, no physical information about the field geometry should be extracted from those solutions.

\input{hotdc_table}

\section{Conclusion} \label{s:conclu}

In this study, we performed the first homogeneous analysis of all known magnetic white dwarfs showing splitting in neutral helium lines reported in the literature.
After a careful examination of all candidates, we could not confirm the magnetic nature of 18 objects and reclassify them as simple DB.
The fact that the average mass of those stars is \Msol{M_*  = 0.61} lends support to the idea that these white dwarfs were misclassified as magnetic.
Assuming an offset dipole geometry for the magnetic field, we could well reproduce the spectra of 8 magnetic DB white dwarfs.
While we were able to reproduce the observed spectra of these few stars to a very high level of detail, there are still some where we cannot reach such precision, even though the presence of stationary components seems to indicate the presence of neutral helium lines.
It is likely that these stars have a much more complex magnetic field structure. 
It is also possible that some stars have a different chemical composition, possibly carbon and oxygen, but the lack of detailed calculation of the line strength and position under very large magnetic field prevent a formal attribution.
We also identify featureless hot objects that are likely highly magnetic white dwarfs with lines spread beyond recognition.
Interestingly, these stars are also among some of the most massive white dwarfs of our sample (\Msol{M_*  > 1.2}).
Although the number of individuals is rather low, magnetic DB white dwarfs also appear to be more massive, on average, than typical DB white dwarfs \citep[][]{Bergeron2011, GenestBeaulieu2019}, suggesting, as has already been proposed by several authors for their hydrogen-rich counterpart, that they could be the result of mergers.

Future work on such objects should consider higher multipole expansion and other chemical composition when the appropriate data become available.
Also, given that rapid rotation is often found among highly magnetic white dwarfs, time resolved spectroscopic observations should be obtained in order to get clues about the exact field geometry.
Finally, it is still unknown how exactly the atmospheric structure and line broadening are affected in the presence of a very large magnetic field.
While neglecting these effects does not prevent us from achieving relatively good fit at moderate fields, those approximations may be responsible in part for our failure to model objects with much higher field strength.
Theoretical calculations exploring those aspects are highly awaited.

\section*{Data Availability}

The data underlying this article are available on the Montreal White Dwarf Database\footnote{\url{https://www.montrealwhitedwarfdatabase.org/}} \citep{MWDD}.

\bibliographystyle{apj}
\bibliography{references}

\appendix

\newcounter{pagecounter}
\newcounter{allfitspages}
\setcounter{allfitspages}{13}

\makeatletter
\@whilenum{\value{pagecounter} < \value{allfitspages}}\do{%
    \stepcounter{pagecounter}
    \centering{
        \includegraphics[page=\value{pagecounter}, width=0.91\linewidth]{figures/fits/all_fits}
        \ifnum \value{pagecounter} = 1 {
                    \phantomsection
                    \label{f:all_fits}
                } \fi
        \clearpage
    }
}
\makeatother



\end{document}

%% file: names.tex

\startlongtable
\begin{deluxetable}{cccrr}
    \tablecaption{Object Names and Coordinates\label{t:names}}
    \tablehead{
        \colhead{J Name} & \colhead{Gaia Source ID} & \colhead{MWDD ID} & \colhead{R.A.} & \colhead{Decl.}
    }
    \startdata
    J0017+0041 & 2545579360599665792 & WD 0015+004                    & 4.42677 & 0.69359 \\
    J0021+1502 & 2792315366213367296 & SDSS J002128.59+150223.8       & 5.36923 & 15.03962 \\
    J0029$-$2133 & 2360911537990561280 & WD 0026$-$218                    & 7.35903 & $-$21.56164 \\
    J0043$-$1000 & 2377863773908424448 & V* BV Cet                      & 10.94093 & $-$10.00755 \\
    J0119+3014 & 310195354299572864 & SDSS J011919.89+301419.6       & 19.83284 & 30.23888 \\
    J0121+3846 & 323521675747536384 & SDSS J012118.42+384606.9       & 20.32687 & 38.76858 \\
    J0124+4003 & 323694161634236160 & SDSS J012426.34+400357.4       & 21.10968 & 40.06593 \\
    J0129$-$3055 & 5016897564123557248 & GD 1363                        & 22.48379 & $-$30.91997 \\
    J0142+1315 & 2587543763765699968 & WD 0140+130                    & 25.68911 & 13.26281 \\
    J0211+0031 & 2507283748560308864 & WD 0208+002                    & 32.81857 & 0.52445 \\
    J0211+2115 & 99498964725981440 & 2MASS J02114816+2115491        & 32.95132 & 21.26327 \\
    J0212+0644 & 2521035817229538688 & PB 6737                        & 33.02212 & 6.73911 \\
    J0251+3515 & 140469482396822912 & SDSS J025150.40+351548.3       & 42.96003 & 35.26337 \\
    J0305+3747 & 142506121529015552 & SDSS J030502.28+374747.2       & 46.25951 & 37.79642 \\
    J0333+0007 & 3264551560189562112 & WD 0330$-$000                    & 53.33526 & 0.12236 \\
    J0343$-$0641 & 3244357341222727168 & WD 0340$-$068                    & 55.78407 & $-$6.69099 \\
    J0543+8340 & 558429448802982272 & SDSS J054326.73+834056.2       & 85.86161 & 83.68228 \\
    J0724+3640 & 897316936986669440 & SDSS J072453.57+364010.6       & 111.22322 & 36.66959 \\
    J0732+1642 & 3169153597742027520 & SDSS J073249.55+164205.5       & 113.20644 & 16.70154 \\
    J0736+6712 & 1096007415499429888 & SDSS J073647.22+671240.5       & 114.19686 & 67.21126 \\
    J0742+3157 & 880354496226790400 & SDSS J074213.37+315703.2       & 115.55557 & 31.95080 \\
    J0800+0655 & 3144237908341731712 & SDSS J080042.47+065542.1       & 120.17677 & 6.92838 \\
    J0822+1201 & 649304840753259520 & SDSS J082247.60+120146.8       & 125.69833 & 12.02964 \\
    J0830+5057 & 1028123052204089472 & SDSS J083047.22+505734.2       & 127.69678 & 50.95952 \\
    J0836+1548 & 657870899622031104 & SDSS J083627.35+154850.3       & 129.11392 & 15.81394 \\
    J0837+5332 & 1030233534709729152 & SDSS J083754.83+533240.3       & 129.47848 & 53.54459 \\
    J0840+5511 & 1031044076641942656 & SDSS J084040.43+551113.4       & 130.16853 & 55.18702 \\
    J0842$-$0222 & 3072348715677121280 & Gaia DR2 3072348715677121280   & 130.56227 & $-$2.37407 \\
    J0844+3129 & 706879014794590976 & GALEX 2737108878850263243      & 131.01865 & 31.48999 \\
    J0847+4842 & 1015028491488955776 & WD 0843+488                    & 131.81740 & 48.70567 \\
    J0849+2857 & 705246450482748288 & SDSS J084929.10+285720.4       & 132.37098 & 28.95563 \\
    J0856+1611 & 611401999180118528 & EGGR 904                       & 134.07880 & 16.18424 \\
    J0856+2534 & 691604221304319744 & GALEX 2693902448450342235      & 134.20675 & 25.57799 \\
    J0910+3941 & 815252481465706752 & SDSS J091004.67+394153.5       & 137.51950 & 39.69828 \\
    J0924+3613 & 798681333703439872 & PM J09249+3613                 & 141.23151 & 36.21871 \\
    J0935+4429 & 815134799361707392 & US 736                         & 143.75874 & 44.49456 \\
    J0938+4740 & 824699760249852544 & SDSS J093832.79+474050.6       & 144.63653 & 47.68060 \\
    J0942+5401 & 1021347930273637888 & SDSS J094209.49+540157.5       & 145.53941 & 54.03265 \\
    J1035+2739 & 728338148874098304 & GALEX 2694887658113466845      & 158.82427 & 27.65950 \\
    J1046$-$0518 & 3776918275016618112 & HE 1043$-$0502                   & 161.54037 & $-$5.30431 \\
    J1121+1039 & 3915674026806527616 & SDSS J112148.80+103934.2       & 170.45311 & 10.65946 \\
    J1142+2052 & 3979498031099140736 & SDSS J114246.46+205222.1       & 175.69354 & 20.87285 \\
    J1144+6629 & 1056998259069523584 & WD 1141+667                    & 176.16316 & 66.49120 \\
    J1153+1331 & 3923217711660623232 & SDSS J115345.97+133106.6       & 178.44141 & 13.51852 \\
    J1155+3148 & 4027359845271139328 & WD 1152+320                    & 178.76523 & 31.80715 \\
    J1202+4034 & 4034928775942285184 & SDSS J120224.38+403455.7       & 180.60152 & 40.58217 \\
    J1214$-$1724 & 3520974164461518592 & HE 1211$-$1707                   & 183.53433 & $-$17.41188 \\
    J1257+1216 & 3737248204724387712 & SDSS J125726.95+121613.4       & 194.36225 & 12.27040 \\
    J1308+8502 & 1726678630833373824 & GJ 3768                        & 197.18267 & 85.04004 \\
    J1317+3917 & 1524545508398868352 & SDSS J131717.75+391719.6       & 199.32390 & 39.28876 \\
    J1328+5908 & 1662221475346089984 & WD 1327+594                    & 202.24176 & 59.14751 \\
    J1341$-$2219 & 6192440730197235072 & HE 1338$-$2204                   & 205.32176 & $-$22.32894 \\
    J1348+1100 & 3727110943738807424 & SDSS J134845.98+110008.8       & 207.19144 & 11.00244 \\
    J1349+2056 & 1250142315600142848 & SDSS J134913.52+205646.9       & 207.30614 & 20.94635 \\
    J1407+3011 & 1453322271887656448 & SDSS J140750.65+301130.2       & 211.96097 & 30.19168 \\
    J1428+1039 & 1177342619932795008 & SDSS J142810.12+103953.7       & 217.04212 & 10.66489 \\
    J1437+3152 & 1286229592893341184 & SDSS J143739.13+315248.8       & 219.41296 & 31.88014 \\
    J1453+0652 & 1160931721694284416 & SDSS J145301.61+065221.0       & 223.25671 & 6.87233 \\
    J1455+1812 & 1188753901361576064 & SDSS J145558.39+181252.4       & 223.99327 & 18.21449 \\
    J1532+1647 & 1207706531182157568 & LSPM J1532+1647                & 233.22902 & 16.79276 \\
    J1532+4914 & 1402291738219385856 & GALEX 2686548949043512447      & 233.10086 & 49.24829 \\
    J1537+2337 & 1220733952969649792 & SDSS J153725.51+233719.6       & 234.35622 & 23.62220 \\
    J1603+5249 & 1404750795975116544 & SDSS J160319.39+524935.9       & 240.83072 & 52.82668 \\
    J1611+0921 & 4453251478106450432 & SDSS J161159.82+092134.1       & 242.99921 & 9.35948 \\
    J1616+1714 & 1198777431614139008 & SDSS J161643.67+171453.2       & 244.18205 & 17.24789 \\
    J1623+0650 & 4439549776517821184 & SDSS J162352.60+065056.8       & 245.96915 & 6.84907 \\
    J1623+3546 & 1329468781009484928 & SDSS J162303.19+354641.0       & 245.76326 & 35.77803 \\
    J1640+5341 & 1425909733315616000 & GD 356                         & 250.23722 & 53.68508 \\
    J1704+3213 & 1310514849813902592 & SDSS J170400.01+321328.7       & 256.00001 & 32.22452 \\
    J1724+3234 & 1333808965722096000 & GALEX 2680391679633002665      & 261.13393 & 32.57089 \\
    J1732+3356 & 4602096559046690816 & SDSS J173232.09+335610.4       & 263.13371 & 33.93619 \\
    J1849+6458 & 2253826832091026560 & Gaia DR2 2253826832091026560   & 282.35944 & 64.96981 \\
    J1900+7039 & 2262849634963004416 & GJ 742                         & 285.04387 & 70.66653 \\
    J2012+3113 & 2053953008490747392 & GD 229                         & 303.09296 & 31.23066 \\
    J2126$-$0024 & 2687624026545008640 & SDSS J212656.71$-$002423.8       & 321.73636 & $-$0.40661 \\
    J2151+0031 & 2681243457490130304 & SDSS J215135.01+003140.2       & 327.89589 & 0.52740 \\
    J2211+2221 & 1782182385825578496 & SDSS J221150.49+222139.6       & 332.96032 & 22.36094 \\
    J2247+1456 & 2732459327587247360 & WD 2245+146                    & 341.92286 & 14.94378 \\
    J2258+2808 & 1883599208067212160 & SDSS J225828.49+280829.0       & 344.61871 & 28.14140 \\
    J2323$-$0046 & 2644029696174113280 & PHL 493                        & 350.90657 & $-$0.77456 \\
    J2346+3853 & 1919346461391649152 & SDSS J234605.44+385337.7       & 356.52261 & 38.89375 \\
    \enddata
    
\end{deluxetable}

%% file: rejected_table.tex

\startlongtable
\begin{deluxetable}{lc}
    \tablecaption{Rejected Candidates\label{t:rejected}}
    \tablehead{
        \colhead{J name} & \colhead{Reason}
    }
    \startdata
    J0021+1502\tablenotemark{a} & Effective temperature too low for DB \\
    J0029$-$2133 & Effective temperature too low for DB \\
    J0121+3846 & Not magnetic \\
    J0124+4003 & Not magnetic \\
    J0129$-$3055 & Effective temperature too low for DB \\
    J0211+0031\tablenotemark{a} & Effective temperature too low for DB \\
    J0211+2115\tablenotemark{a} & Not magnetic \\
    J0251+3515 & Not magnetic \\
    J0333+0007\tablenotemark{a} & Effective temperature too low for DB \\
    J0343$-$0641\tablenotemark{a} & Shallow hydrogen features \\
    J0543+8340 & Not magnetic \\
    J0724+3640 & Not magnetic \\
    J0736+6712 & Too noisy \\
    J0742+3157\tablenotemark{a} & Effective temperature too low for DB \\
    J0800+0655\tablenotemark{a} & Effective temperature too low for DB \\
    J0840+5511 & Not magnetic \\
    J0844+3129 & Not magnetic \\
    J0847+4842\tablenotemark{a} & DBA (Balmer lines visible) \\
    J0849+2857\tablenotemark{a} & Effective temperature too low for DB \\
    J0856+2534\tablenotemark{a} & Effective temperature too low for DB \\
    J0910+3941 & Too noisy \\
    J0924+3613\tablenotemark{a} & Effective temperature too low for DB \\
    J0938+4740 & Effective temperature too low for DB \\
    J1035+2739 & Not magnetic \\
    J1121+1039\tablenotemark{a} & Effective temperature too low for DB \\
    J1142+2052 & Effective temperature too low for DB \\
    J1144+6629\tablenotemark{a} & Effective temperature too low for DB \\
    J1153+1331\tablenotemark{a} & Too noisy \\
    J1155+3148 & Not magnetic \\
    J1202+4034\tablenotemark{a} & Too noisy \\
    J1308+8502\tablenotemark{a} & Effective temperature too low for DB \\
    J1317+3917 & Not magnetic \\
    J1328+5908\tablenotemark{a} & Possibly double magnetic DA binary \\
    J1341$-$2219 & Possibly DBA \\
    J1349+2056\tablenotemark{a} & Effective temperature too low for DB \\
    J1428+1039 & Not magnetic \\
    J1532+1647\tablenotemark{a} & Possibly DQ \\
    J1532+4914 & Not magnetic \\
    J1537+2337 & Not magnetic \\
    J1616+1714 & Not magnetic \\
    J1623+0650\tablenotemark{a} & Not a DB \\
    J1623+3546\tablenotemark{a} & Too noisy \\
    J1640+5341\tablenotemark{a} & Effective temperature too low for DB \\
    J1704+3213\tablenotemark{a} & Not a DB \\
    J1732+3356 & Not magnetic \\
    J1849+6458 & Effective temperature too low for DB \\
    J2126$-$0024 & Not magnetic \\
    J2151+0031\tablenotemark{a} & Effective temperature too low for DB \\
    J2211+2221 & Not magnetic \\
    J2258+2808\tablenotemark{a} & Effective temperature too low for DB \\
    \enddata
    \tablenotetext{a}{Labeled as Magnetic non DA in \citet{Hardy2022}}
\end{deluxetable}

%% file: average_mass.tex
$0.68 \, M_\odot$ ($0.83 \, M_\odot$ if we exclude the stars that hit the lower limit)%

%% file: figures/mwd_db.tex
\def \dbfigpath {./figures/fits}

\begin{figure}
\centering
\includegraphics[width=\linewidth]{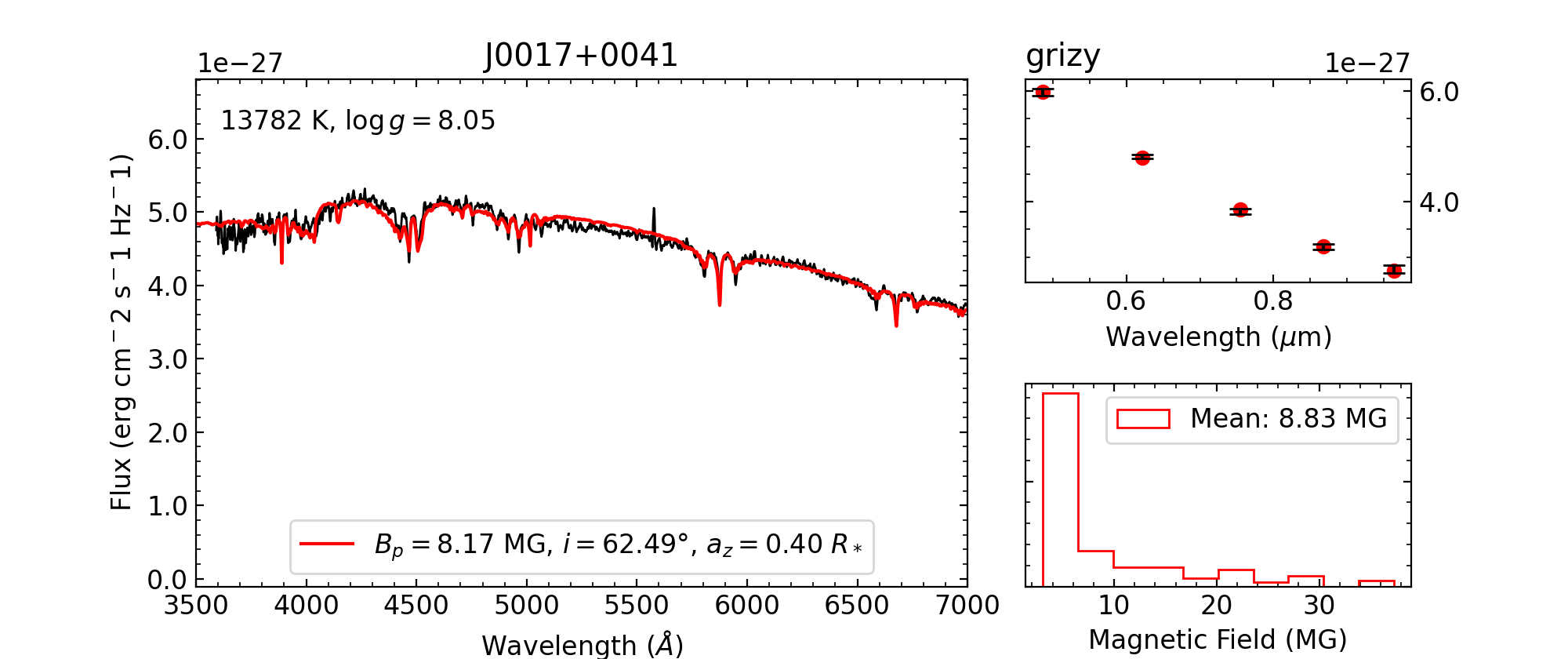}
\caption{\label{f:good_fit_0} Dipole solution for star J0017+0041. The left panel is our best fit to the neutral helium lines. Upper right panel represents the best fit to the $ugriz$ photometry while the bottom right panel shows the distribution of magnetic field elements at the visible surface of the star for our best fit solution.}
\end{figure}

\begin{figure}
\centering
\includegraphics[width=\linewidth]{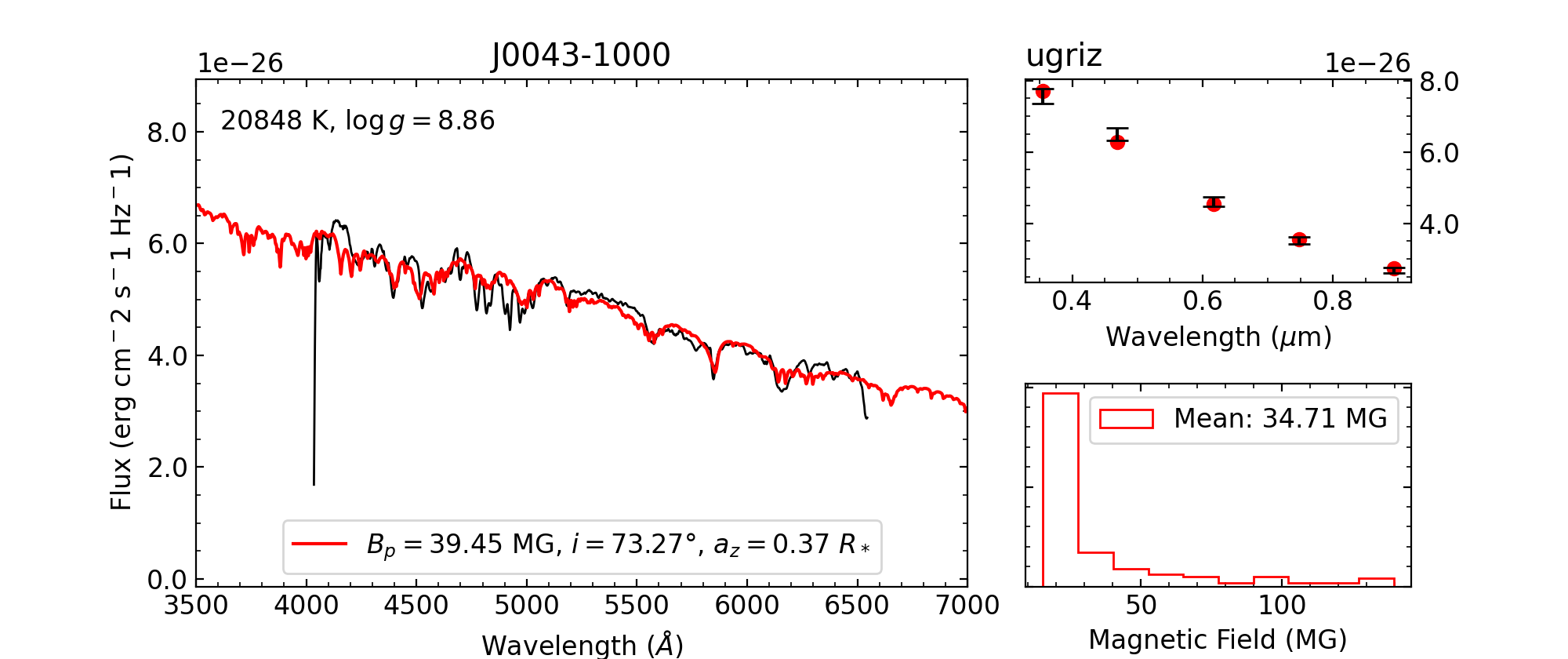}
\caption{\label{f:good_fit_1} Dipole solution for star J0043-1000.}
\end{figure}

\begin{figure}
\centering
\includegraphics[width=\linewidth]{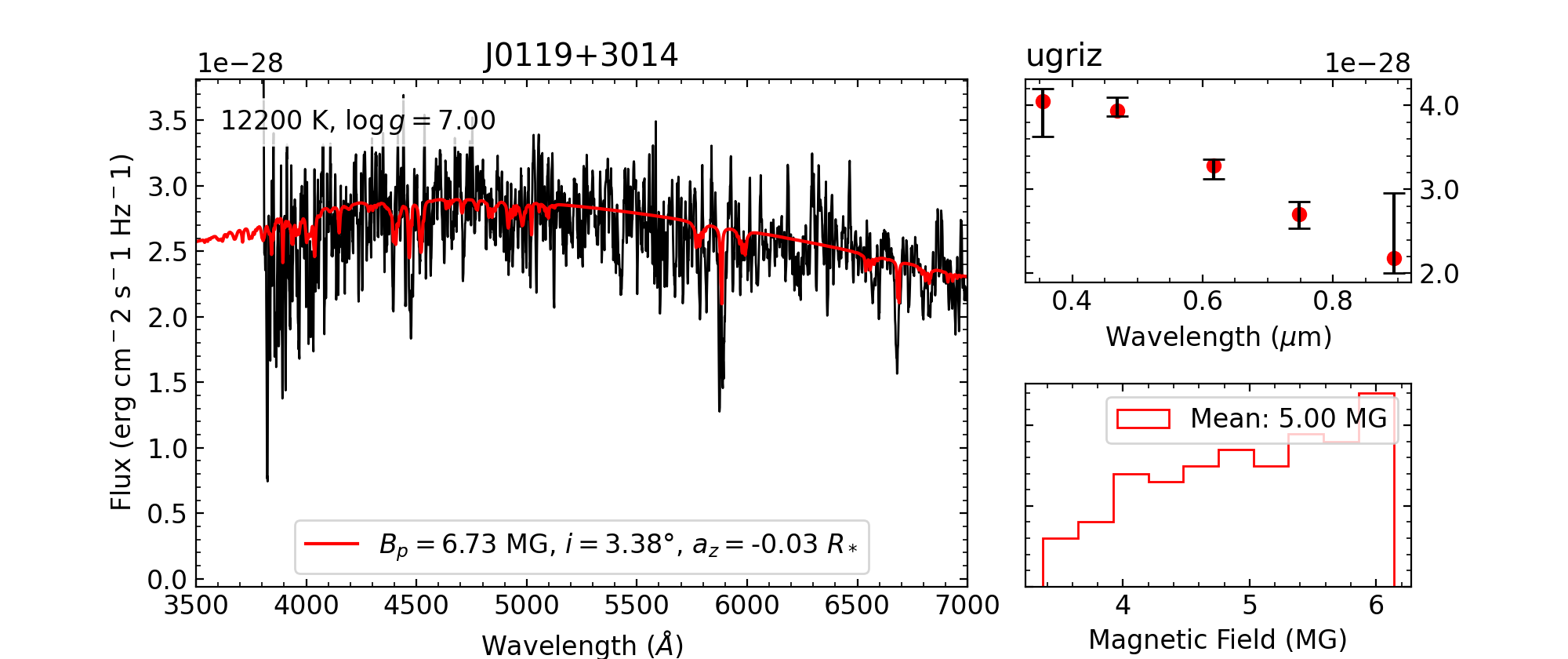}
\caption{\label{f:good_fit_2} Dipole solution for star J0119+3014.}
\end{figure}

\begin{figure}
\centering
\includegraphics[width=\linewidth]{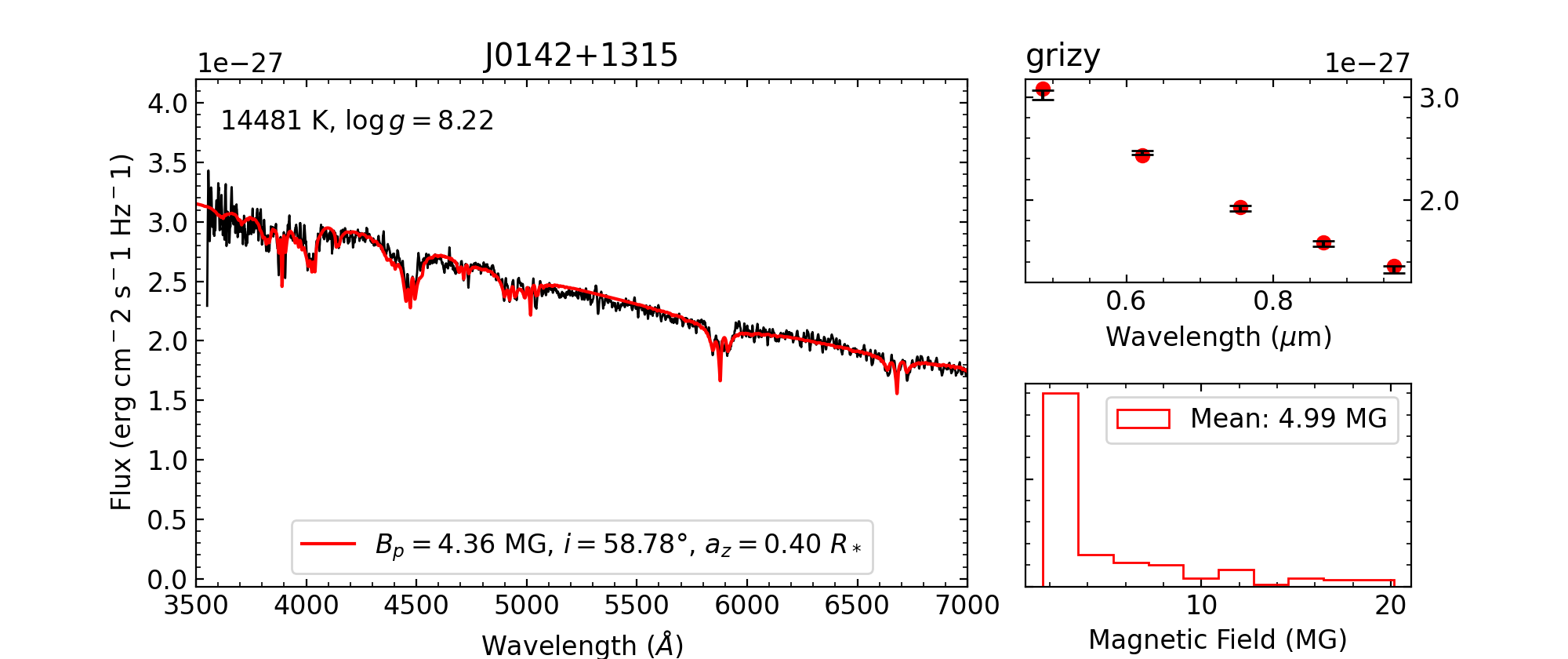}
\caption{\label{f:good_fit_3} Dipole solution for star J0142+1315.}
\end{figure}

\begin{figure}
\centering
\includegraphics[width=\linewidth]{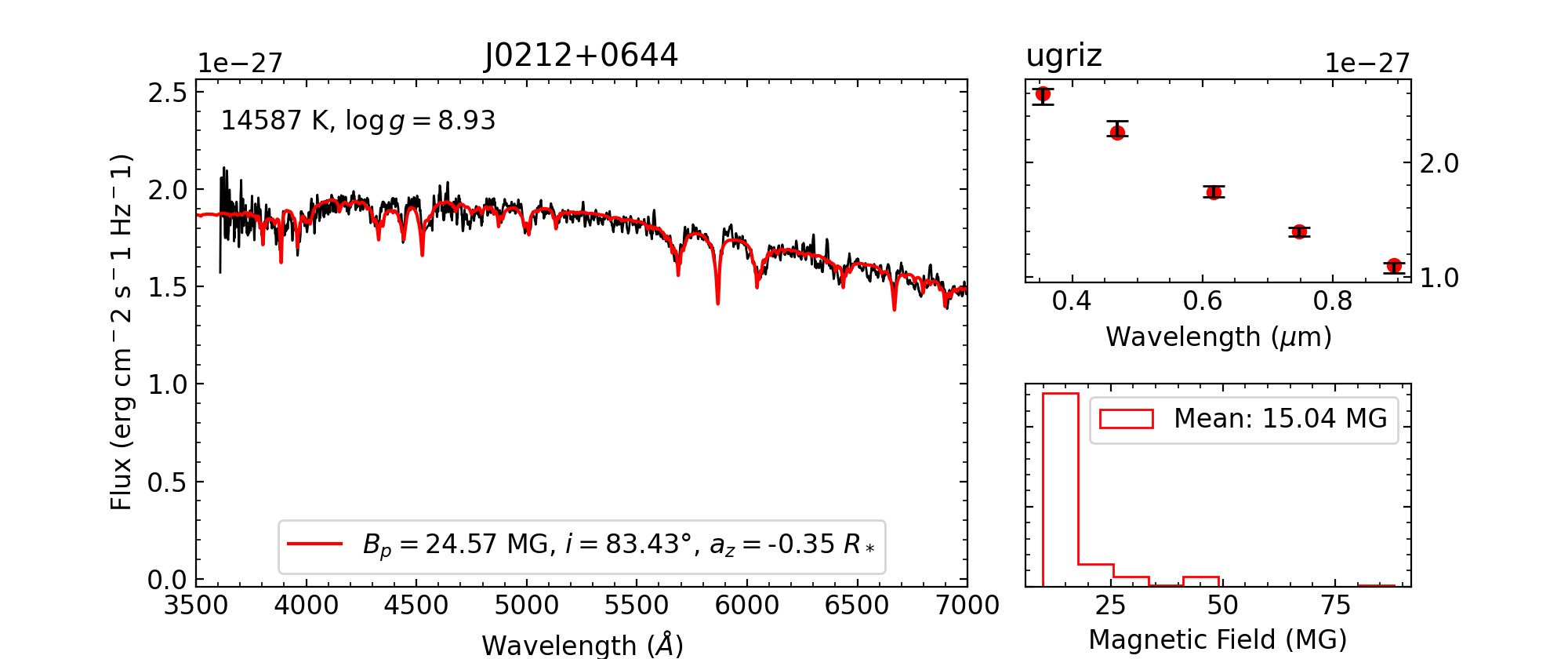}
\caption{\label{f:good_fit_4} Dipole solution for star J0212+0644.}
\end{figure}

\begin{figure}
\centering
\includegraphics[width=\linewidth]{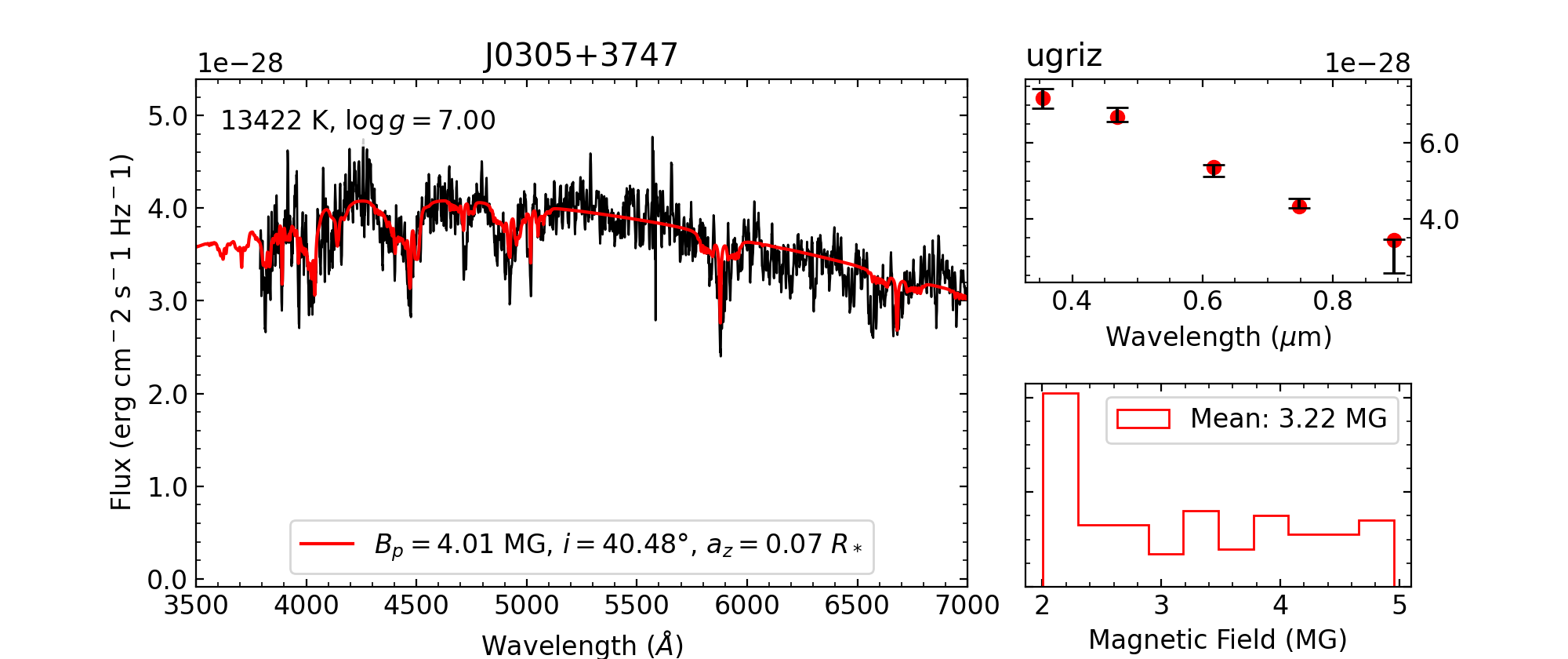}
\caption{\label{f:good_fit_5} Dipole solution for star J0305+3747.}
\end{figure}

\begin{figure}
\centering
\includegraphics[width=\linewidth]{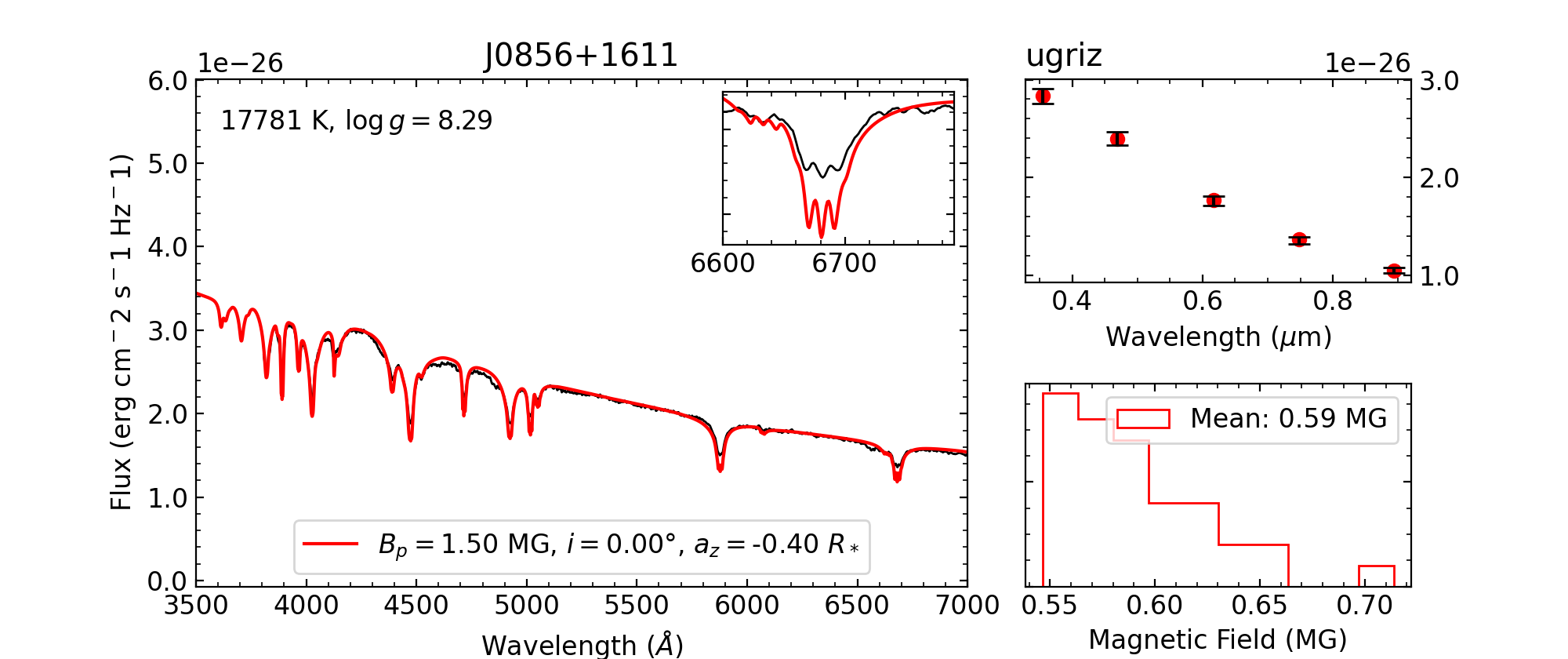}
\caption{\label{f:good_fit_6} Dipole solution for star J0856+1611.}
\end{figure}

\begin{figure}
\centering
\includegraphics[width=\linewidth]{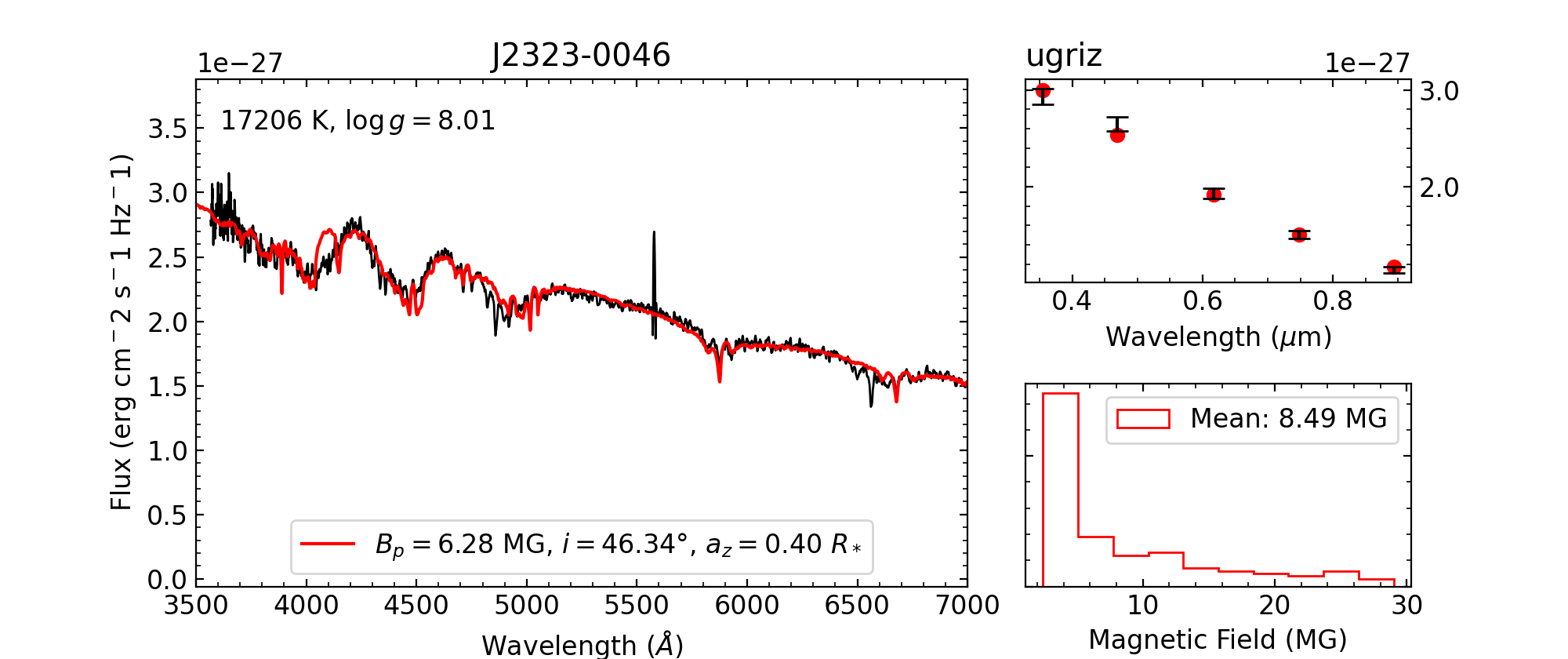}
\caption{\label{f:good_fit_7} Dipole solution for star J2323-0046.}
\end{figure}

%% file: results_table.tex

\begin{deluxetable}{lccccccc}
    \tablecaption{\label{t:results} Model fits with offset dipole geometry}
    \tablehead{
        \colhead{J name} & \colhead{$T_{\rm{eff}}$ (K)} & \colhead{$\log{g}$} & \colhead{Mass ($M_\odot$)} & \colhead{$B_p$ (MG)} & \colhead{$i$ (deg)} & \colhead{$a_z$ ($R_*$)} & \colhead{Mean $B$ (MG)}
    }
    \startdata
    J0017+0041 & 13782 (207) & 8.05 (0.04) & 0.62 (0.05) &   8.17 (0.12) & 62 (1) &  0.40 (0.00) & 8.83 \\
    J0043$-$1000 & 20848 (1077) & 8.86 (0.18) & 1.13 (0.18) &  39.45 (0.46) & 73 (1) &  0.37 (0.01) & 34.71 \\
    J0119+3014 & 12200 (410) & $< 7.00$ & \textemdash &   6.73 (1.43) &  3 (6) & $-$0.03 (0.10) & 5.00 \\
    J0142+1315 & 14481 (652) & 8.22 (0.05) & 0.72 (0.07) &   4.36 (0.08) & 59 (1) &  0.40 (0.00) & 4.99 \\
    J0212+0644\tablenotemark{a} & 14591 (251) & 8.93 (0.07) & 1.16 (0.07) &  24.57 (0.61) & 83 (3) & $-$0.35 (0.01) & 15.04 \\
    J0305+3747 & 13422 (489) & $< 7.00$ & \textemdash &   4.01 (0.03) & 40 (3) &  0.07 (0.01) & 3.22 \\
    J0856+1611 & 17781 (219) & 8.29 (0.03) & 0.78 (0.03) &   1.50 (0.00) &  0 (0) & $-$0.40 (0.00) & 0.59 \\
    J2323$-$0046 & 17206 (1146) & 8.01 (0.13) & 0.60 (0.15) &   6.28 (0.13) & 46 (1) &  0.40 (0.00) & 8.49 \\
    \enddata
    \tablenotetext{a}{Labeled as Magnetic non DA in \citet{Hardy2022}}
\end{deluxetable}

%% file: magnetic_non_DB_table.tex

\begin{deluxetable}{lll}
    \tablecaption{Magnetic white dwarfs we could not fit within our theoretical framework\label{t:nonDB}}
    \tablehead{
        \colhead{} & \colhead{J name} & \colhead{}
    }
    \startdata
    J0822+1201\tablenotemark{a} & J0830+5057\tablenotemark{a} & J0836+1548\tablenotemark{a} \\
    J0842$-$0222 & J1046$-$0518 & J1214$-$1724 \\
    J1257+1216\tablenotemark{a} & J1348+1100\tablenotemark{a} & J1407+3011\tablenotemark{a} \\
    J1453+0652\tablenotemark{a} & J1455+1812\tablenotemark{a} & J1724+3234\tablenotemark{a} \\
    J1900+7039\tablenotemark{a} & J2012+3113 & J2247+1456\tablenotemark{a} \\
    \enddata
    \tablenotetext{a}{Labeled as Magnetic non DA in \citet{Hardy2022}}
\end{deluxetable}

%% file: hotdc_table.tex

\begin{deluxetable}{lccc}
    \tablecaption{\label{t:hotdc} Atmospheric parameters of stars classified as Hot DC}
    \tablehead{
        \colhead{J name} & \colhead{$T_{\rm{eff}}$ (K)} & \colhead{$\log{g}$} & \colhead{Mass ($M_\odot$)}
    }
    \startdata
    J0732+1642\tablenotemark{a} & $> 30000$ & \textemdash & \textemdash \\
    J0837+5332 & $> 30000$ & \textemdash & \textemdash \\
    J0935+4429\tablenotemark{a} & 29579 (1525) & 9.03 (0.08) & 1.20 (0.08) \\
    J0942+5401 & 21169 (4181) & 9.04 (0.20) & 1.21 (0.17) \\
    J1437+3152 & 27607 (1415) & 9.17 (0.68) & 1.26 (0.78) \\
    J1603+5249 & 22093 (1290) & 9.11 (0.72) & 1.24 (0.70) \\
    J1611+0921 & $> 30000$ & \textemdash & \textemdash \\
    J2346+3853\tablenotemark{a} & 29503 (1047) & 9.04 (0.35) & 1.21 (0.31) \\
    \enddata
    \tablenotetext{a}{Labeled as Magnetic non DA in \citet{Hardy2022}}
\end{deluxetable}